\documentclass{article} 

\usepackage{amsmath,amsfonts,bm}

\def\1{\bm{1}}

\DeclareMathAlphabet{\mathsfit}{\encodingdefault}{\sfdefault}{m}{sl}
\SetMathAlphabet{\mathsfit}{bold}{\encodingdefault}{\sfdefault}{bx}{n}

\newcommand{\E}{\mathbb{E}}

\newcommand{\R}{\mathbb{R}}

\DeclareMathOperator*{\argmin}{arg\,min}

\usepackage[hidelinks]{hyperref}
\usepackage{url}
\usepackage{graphicx}
\usepackage{amssymb}
\usepackage{algorithm}
\usepackage{algpseudocode}
\usepackage{multirow}
\usepackage{adjustbox}
\usepackage{tabularx,booktabs}
\usepackage{mathtools}
\usepackage{amsthm}
\usepackage{wrapfig}
\usepackage{caption}
\usepackage{xcolor,colortbl}
\usepackage[numbers,sort,compress]{natbib}
\usepackage{fullpage}

\definecolor{lightgreen}{rgb}{.9,1,.9}
\definecolor{darkblue}{rgb}{0,0 ,0.542}
\definecolor{lightgreen}{rgb}{.9,1,.9}
\definecolor{lightblue}{rgb}{.9,.9,1.}
\newcommand{\boxsize}{53}

\hypersetup{
    colorlinks = true,
    citecolor = darkblue,
    linkcolor = {black}
}

\DeclareMathOperator{\reschedule}{reschedule}

\theoremstyle{plain}
\newtheorem*{theorem*}{Theorem}

\def\log{{\mathsf{log}}}

\def\R{\mathbb{R}}
\def\E{\mathbb{E}}

\def\ebm{{\bm{e}}}

\def\xbm{{\bm{x}}}
\def\zbm{{\bm{z}}}
\def\ybm{{\bm{y}}}
\def\zbm{{\bm{z}}}

\def\Abm{{\bm{A}}}
\def\Dbm{{\bm{D}}}
\def\Pbm{{\bm{P}}}
\def\Fbm{{\bm{F}}}
\def\Sbm{{\bm{S}}}

\def\Ibm{{\bm{I}}}

\def\thetabm{{\bm{\theta }}}

\def\Ibf{{\mathbf{I}}}

\def\Abm{{\bm{A}}}
\def\Dbm{{\bm{D}}}
\def\Pbm{{\bm{P}}}
\def\Fbm{{\bm{F}}}
\def\Ibm{{\bm{I}}}

\def\Ncal{{\mathcal{N}}}

\def\Rsf{{\mathsf{R}}}

\def\Hsf{{\mathsf{H}}}

\def\Hsf{{\mathsf{H}}}

\def\argmin{\mathop{\mathsf{arg\,min}}} 

\newcommand{\norm}[1]{\left\lVert#1\right\rVert}

\title{ADOBI: Adaptive Diffusion Bridge For Blind Inverse Problems\\ with Application to MRI Reconstruction}

\author{Yuyang Hu$^{\footnotesize *}$, Albert Peng$^{\footnotesize *}$,  Weijie Gan, and Ulugbek S. Kamilov\\
Washington University in St.~Louis, MO, USA\\
$^{\footnotesize *}$\small These authors contributed equally.\\
\texttt{\{h.yuyang, albertpeng, weijie.gan, kamilov\}@wustl.edu}
}

\begin{document}

\maketitle

\begin{abstract}
Diffusion bridges (DB) have emerged as a promising alternative to diffusion models for imaging inverse problems, achieving faster sampling by directly bridging low- and high-quality image distributions. While incorporating measurement consistency has been shown to improve performance, existing DB methods fail to maintain this consistency in blind inverse problems, where the forward model is unknown. To address this limitation, we introduce ADOBI (Adaptive Diffusion Bridge for Inverse Problems), a novel framework that adaptively calibrates the unknown forward model to enforce measurement consistency throughout sampling iterations. Our adaptation strategy allows ADOBI to achieve high-quality parallel magnetic resonance imaging (PMRI) reconstruction in only 5–10 steps. Our numerical results show that ADOBI consistently delivers state-of-the-art performance, and further advances the Pareto frontier for the perception-distortion trade-off.
\end{abstract}

\section{Introduction}
Many imaging problems can be formulated as inverse problems that seek to recover an unknown image $\xbm\in\R^n$ from from its corrupted observation 
\begin{equation}
\label{Eq:InverseProblem}
\ybm=\Abm\xbm + \ebm, 
\end{equation}
where $\Abm\in\R^{m\times n}$ is a measurement operator, $\ebm\in\R^n$ is the noise, and $\ybm\in\R^n$ is the noisy measurement. In many applications, these problems are further complicated by incomplete knowledge of the measurement operator $\Abm$, resulting in a \emph{blind inverse problem} that requires simultaneous estimation of both the image and the operator. Accelerated parallel magnetic resonance imaging (PMRI) is a prototypical blind inverse problem requiring the recovery of an unknown image and unknown coil sensitivity maps (CSMs) from undersampled and noisy k-space measurements~\cite{Akcakaya.etal2022}.

DL has emerged as a powerful tool for addressing inverse problems~\cite{McCann.etal2017, Ongie.etal2020, Wen.etal2023}. Instead of explicitly defining a regularizer, DL methods use
deep neural networks (DNNs) to map the measurements to the desired images~\cite{Wang2016.etal, DJin.etal2017, Kang.etal2017, Chen.etal2017, delbracio2021projected}. Model-based DL (MBDL) is a widely-used sub-family of DL algorithms that integrate physical measurement models with priors specified using CNNs (see reviews by~\cite{Ongie.etal2020, Monga.etal2021}). The
literature of MBDL is vast, but some well-known examples include plug-and-play priors (PnP), regularization by denoising (RED), deep unfolding (DU), compressed sensing using generative models (CSGM), and deep equilibrium models (DEQ)~\cite{Bora.etal2017, Kamilov.etal2022, Romano.etal2017, zhang2018ista, Hauptmann.etal2018, Gilton.etal2021, Monga.etal2021, hurestoration}. More recent developments have further improved performance of MBDL for blind inverse problems by designing models that jointly estimate both the image and the unknown measurement operator~\cite{arvinteDeep2021, junJoint2021, sriramEndtoend2020}.

Diffusion models (DMs)~\cite{croitoru2023diffusion, ho2020denoising, song2021scorebased} have emerged as powerful tools for addressing inverse problems, due to their ability to model complex data distributions. Once trained, DMs can use their learned score functions to guide posterior sampling, enabling a denoising process that produces reconstructions consistent with observed measurements. This approach enables DMs to tackle highly ill-posed inverse problems, achieving impressive perceptual quality~\cite{whang2022deblurring, ren2023multiscale, xie2022measurement, chung2022come, chung2023diffusion, zhu2023denoising, wang2023zeroshot, feng2023score, wu2024principled, song2024solving}. To handle blind inverse problems, several DM-based variants simultaneously sample both the reconstructed image and the unknown forward operator~\cite{chung2023parallel, murata2023gibbsddrm, bai2024blind}. However, DMs are known to have slow inference due to the need to run many iterations to generate an image from pure noise.

\begin{figure*}[t]
    \centering
    \includegraphics[width=\linewidth]{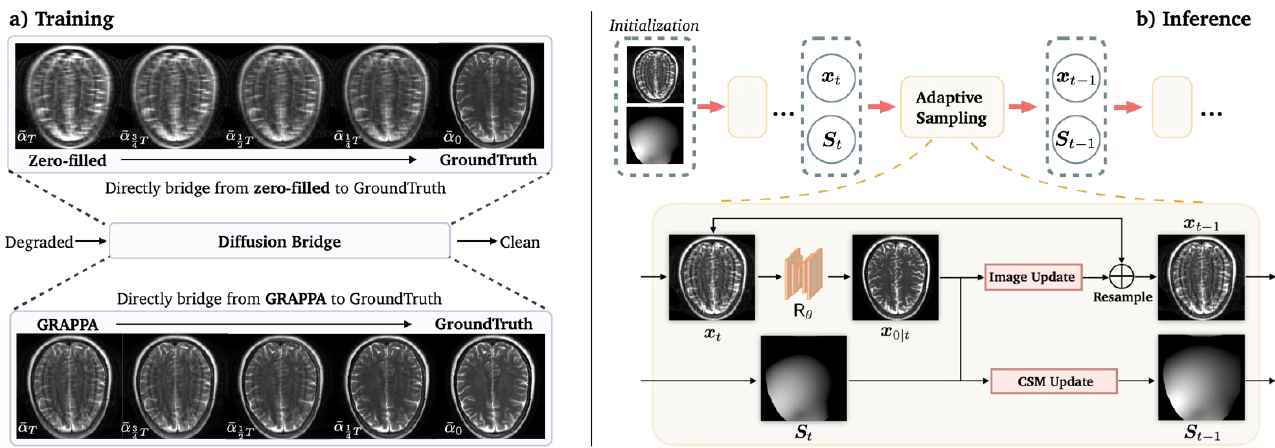}
    \caption{A schematic diagram of the ADOBI training and inference for the PMRI reconstruction. ADOBI begins by initializing the image with either zero-filling or GRAPPA~\cite{griswoldGeneralized2002} and initializes the coil sensitivity maps (CSMs) using ESPIRiT~\cite{ueckerESPIRiTan2014}. It then performs alternating updates, iteratively refining both the image and the measurement operator to ensure consistency with the observed measurements.}
\end{figure*}

Diffusion bridges (DBs)~\cite{liu2023i2sb, indi2023, shi2024diffusion} have been shown to significantly reduce the number of inference steps by using a diffusion process that directly maps corrupted observations to the clean target. Consistent Direct Diffusion Bridge (CDDB)~\cite{chung2024direct} further improves DBs by incorporating measurement-consistent guidance, ensuring data fidelity to observed measurements in non-blind inverse problems where the forward model is fully known. However, current DB methods, including CDDB, cannot enforce data consistency in blind inverse problems, where the forward model is unknown and must be estimated simultaneously with the unknown image. This lack of data consistency constraints in blind settings prevents existing DB approaches from ensuring that reconstructed images accurately align with observed measurements—a crucial requirement for practical applications such as PMRI.

We propose ADOBI (Adaptive Diffusion Bridge for Inverse Problems), a novel framework designed to solve blind inverse problems with measurement consistency. ADOBI uses a pre-trained DB backbone, eliminating the need for retraining or using an extra network to calibrate the forward model.  Our work offers three main contributions:
\begin{enumerate}
    \item \textbf{First measurement-consistent DB method for blind inverse problems:} Unlike CDDB~\cite{chung2024direct}, which requires the exact knowledge of the forward model, ADOBI is the first measurement-consistent DB method specifically designed for blind inverse problems. With our adaptation strategy, ADOBI enables simultaneous estimation of the unknown forward model and the unknown image.
    
    \item \textbf{First image domain DB method for PMRI:} In contrast to previous Fourier-constrained diffusion bridge (FDB)~\cite{mirza2023learning} for PMRI, which operates in the frequency domain, ADOBI is designed for image-domain diffusion bridges. This provides ADOBI with compatibility to leverage existing frameworks and initialization strategies, leading to improvements in reconstruction quality.
    
    \item  \textbf{State-of-the-art performance:} ADOBI achieves state-of-the-art performance on the fastMRI dataset, delivering the best balance between the distortion and perceptual qualities. ADOBI enhances stability compared to conventional DM methods and provides uncertainty quantification, offering a reliable and interpretable method for PMRI reconstruction.
    
\end{enumerate}

\section{Background}

\subsection{Parallel MRI reconstruction}
Consider the following PMRI measurement model~\cite{lustigSparse2007}
\begin{equation}
\label{equ:inverse_problem}
   \ybm = \underbrace{\Pbm\Fbm\Sbm}_{\Abm}\xbm + \ebm\ ,
\end{equation}
where $\Pbm$ is the k-space subsampling pattern, $\Fbm$ is the Fourier transform operator, $\Sbm = ( \Sbm_1,\cdots,\Sbm_{n_c})$ are the CSMs, and $\ebm$ is the noise vector. In this forward model, only the CSMs $\Sbm$ are unknown, making them a critical factor in accurately reconstructing the image $\xbm$.

When CSMs are assumed to be known, image reconstruction can be approached as a regularized optimization problem~\cite{Fessler2020, Liu.etal2022b}, where $f(\xbm) = g(\xbm) + h(\xbm)$, with $g$ acting as the data fidelity term to ensure consistency with the observed data $\ybm$, and $h$ serving as a regularizer that incorporates prior information about $\xbm$. Common choices for these terms in CS-MRI include the least-squares data fidelity term, $g(\xbm) = \frac{1}{2}\norm{\Abm\xbm - \ybm}_2^2$, and the total variation (TV) regularizer, $h(\xbm) = \tau\norm{\Dbm\xbm}_1$, where $\Dbm$ denotes the image gradient, and $\tau > 0$ is a parameter controlling the trade-off between data fidelity and image smoothness.

In practice, exact knowledge of the coil sensitivity maps (CSMs) is rarely available; instead, CSMs are typically estimated from the auto-calibration signal (ACS) region in undersampled k-space data~\cite{ueckerESPIRiTan2014}. Estimating CSMs from a limited set of ACS lines, however, can introduce inaccuracies that degrade the quality of the final reconstruction. Recent methods demonstrate that jointly estimating the image and CSMs can significantly improve reconstruction performance~\cite{arvinteDeep2021, junJoint2021, sriramEndtoend2020, Gan_2021_ICCV, hu2024spicer, gan2023block}.

\subsection{Diffusion models for imaging inverse problems}

 Diffusion models (DMs) are designed to learn a parameterized Markov chain to convert the Gaussian distribution to the target data distribution $p(\xbm)$ (see~\cite{Daras.etal2024} for a recent review). The forward diffusion process in the now classical DDPM method~\cite{ho2020denoising} gradually degrades a clean image $\xbm_0$ with the Gaussian noise to obtain the transition probability:
\begin{equation}
    q(\xbm_t|\xbm_{t-1}) := \Ncal{(\xbm_t; \sqrt{1-\beta_t}\xbm_{t-1},\beta_t\Ibf}),
\end{equation}
where $t\in[0, T]$, $\Ncal(\cdot)$ denotes the Gaussian probability density function (pdf), $\beta_{t}$ denotes to the noise schedule of the process. Using $\alpha_t := 1-\beta_{t}$ and $\bar{\alpha}_t := \prod^{t}_{s=1}\alpha_s$, one can rewrite $\xbm_t$ as a linear combination of noise $\bm\epsilon$ and $\xbm_0$
\begin{equation}
    \xbm_t = \sqrt{\bar{\alpha}_t}\xbm_0 + \sqrt{1-\bar{\alpha}_t}\bm\epsilon,
\end{equation}
where $\epsilon \sim \Ncal(0, \Ibf)$. This allows a closed-form expression
for the marginal distribution for sampling $\xbm_t$ given $\xbm_0$
\begin{equation}
    q(\xbm_t|\xbm_{0}) := \Ncal(\xbm_t; \sqrt{\bar{\alpha}_t}\xbm_0, (1-\bar{\alpha}_t)\Ibf).
\end{equation}
The objective of the reverse diffusion process is to sample a clean image from the learned distribution. This is achieved by training a deep neural network (DNN) to reverse the Markov Chain from $\xbm_T$ to $\xbm_0$. The DNN $\bm{\epsilon}_\thetabm$ is learned by minimizing the following expected loss
\begin{equation}
\label{Eq:dm_training}
\mathcal{L}(\thetabm) = \E_{\bm\epsilon, \xbm_t, t} [ \norm{\bm\epsilon_\thetabm(\xbm_t, t) - \bm\epsilon}_2^2].
\end{equation} 
DDPM inference uses the pre-trained DNN $\bm\epsilon_\thetabm$ to approximate the score, leading to the following sampling process:
\begin{equation}
    \xbm_{t-1} = \frac{1}{\sqrt{\alpha_t}}(\xbm_t - \frac{\beta_t}{\sqrt{1-\bar{\alpha}_t}}\bm\epsilon_\thetabm(\xbm_t,t)) + \sigma_t\zbm,
\label{eq:ddpm_sampling}
\end{equation} 
 where $\zbm \sim  \Ncal(0, \Ibf)$. This reparameterization can be viewed as gradual noise removal, ultimately producing a clean sample from the learned data distribution.
 
A pre-trained denoising diffusion model provides an approximation of the score function $\nabla \log p_{\sigma}(\xbm)$ at each noise level~\cite{efron2011tweedie}. The score can be used to guide image reconstruction in inverse problems. A common strategy is to integrate data consistency updates into the posterior sampling process~\cite{chung2023diffusion, zhu2023denoising, wang2023zeroshot, feng2023score, wu2024principled, song2024solving}. Here, after each reverse diffusion step, a data consistency step ensures that generated samples respect the specific measurement constraints associated with the inverse problem. One popular example is DPS~\cite{chung2023diffusion}, which imposes a measurement constrained posterior sampling step after DDPM sampling in~(\ref{eq:ddpm_sampling}) :
\begin{equation}
    \hat{\xbm}_{t-1} = \xbm_{t-1} + \gamma \Abm^\Hsf(\ybm - \Abm\xbm_{0|t}).
\end{equation}
This measurement constrained update enables sampling from the conditional distribution $p(\xbm|\ybm)$.

 DM-based posterior sampling was recently extended to blind inverse problems, where both the image and the unknown forward model require simultaneous estimation. These methods incorporate additional priors to jointly optimize the image and forward model parameters, enabling recovery of both. This joint optimization framework has been successfully demonstrated in blind deblurring and imaging through turbulence~\cite{chung2023parallel, sanghvi2023kernel, bai2024blind, murata2023gibbsddrm}, broadening the practical impact and applicability of DMs in real-world scenarios.

\subsection{Diffusion bridge for imaging inverse problems}

The Diffusion Bridge (DB) extends the Diffusion Model (DM) framework by specifing a stochastic process that connects two paired distributions, without restricting the initial distribution to the random Gaussian. In imaging inverse problems, DB constructs a probabilistic path between two states (e.g., degraded and clean images) using the Schrödinger bridge formulation, which has demonstrated improvements in both performance and inference speed over the DM approaches~\cite{indi2023, liu2023i2sb, chung2024direct, hu2024stochastic}. The Direct Diffusion Bridge~\cite{chung2024direct} brings these frameworks together into a unified approach, showing that they are effectively equivalent despite originating from different theoretical ideas.

Under the unified Direct Diffusion Bridge (DDB) framework, the transition from a degraded image to a clean image can be defined as follows. Let $ \zbm $ denote the initial degraded image from source distribution $p(\zbm|\xbm)$ and $ \xbm $ the target clean image from distribution $p(\xbm)$. The forward diffusion process starts at $ \zbm $ and progressively transforms it to $ \xbm $, with each intermediate state $ \xbm_t $ representing a convex mixture of both. The forward process is defined as:
\begin{equation}
  \xbm_t = (1 - \alpha_t)\xbm_0 + \alpha_t \xbm_1 + \sigma_t \bm\epsilon, \quad \bm\epsilon \sim \Ncal(0, \Ibf),
  \label{eq:DB}
\end{equation}
where $\{\alpha_t, \sigma_t \}$ is the signal/noise schedule governing the evolution of the diffusion process over time. Once this forward process for DB is constructed, a neural network is trained to predict the posterior mean at each step by minimizing the object function:
\begin{equation}
\E_{(\xbm, \ybm) \sim p(\xbm, \ybm), t \sim p(t)}\left\| \Rsf_{\thetabm} \left(\xbm_t, t \right) - \xbm_0 \right\|^2_2.
\end{equation}
The network is trained  to perform Minimum Mean Square Error (MMSE) estimation of the target image at each step, conditioned on the degraded input. By reparametrizing the process, it becomes clear that estimating the target image is analogous to predicting an intermediate state of the DB process at $t-1$~\cite{indi2023}, which can be expressed as
\begin{equation}
\begin{matrix}
\xbm_{t-1} = (1 - \beta_{t})\xbm_{0|t} + \beta_{t} \xbm_t +\left(\sigma_t\beta_{t}-\sigma_{t-1} \right)\bm{\epsilon}, 
\end{matrix}
\label{eq:DDB}
\end{equation}
where $\beta_{t} = ({\alpha}_{t-1}/\alpha_t)$.
This equation shows that the estimate at the next time step $t-1$ is influenced by both the current prediction of the target, $\xbm_{0|t} = \E[\xbm_{0}|\xbm_{t}]$, and the degraded image $\xbm_t$ at time $t$. With this approach, DDB is able to recover the high-quality image from the low-quality estimation, refining the reconstruction at each step.

While DDB achieves impressive restoration performance, it lacks an explicit mechanism to ensure that the reconstructed image aligns precisely with measurement data. The Consistent Direct Diffusion Bridge (CDDB)~\cite{chung2024direct} addresses this by enforcing a data-consistency constraint after each sampling step~(\ref{eq:DDB}), which can be written as
\begin{equation}
\begin{matrix}
\hat{\xbm}_{t-1} = \xbm_{t-1} + \gamma \Abm^\Hsf(\ybm - \Abm\xbm_{0|t}),
\end{matrix}
\label{eq:CDDB}
\end{equation} 
Data-consistency constraint improves the fit between the DB estimate $\xbm_s$ and the measurement $\ybm$, thus leading to a more accurate and reliable reconstruction.

Despite its potential, the data-consistency constraint of CDDB relies on a fully known measurement operator, limiting its applicability in blind inverse problems. Moreover, CDDB has yet to be explored for the PMRI reconstruction task, where high fidelity to measurements is critical, and the measurement model is generally unknown.

\begin{algorithm}[t]
  \caption{ADOBI Training} 
  \begin{algorithmic}[1]
     \Require  \text{paired dataset (image, measurement):} $(\xbm,\ybm)$
    \State \textbf{repeat:} 
    \State \quad $\xbm, \ybm \sim p(\xbm, \ybm)$, $\alpha_t \sim p(\alpha_t) $
    \State \quad $\zbm \gets \mathsf{Initialization}(\ybm)$
    \State \quad  $\xbm_t = (1 - \alpha_t)\xbm + \alpha_t \zbm + \sigma_t \bm{\epsilon}$
    \State \quad Reduce the loss function \par   \quad $\left\| \Rsf_\theta \left(\xbm_t, t \right) - \xbm \right\|^2_2$ 
    \State \textbf{until converge} 
\end{algorithmic}
    \label{alg:training}   
\end{algorithm}

\begin{algorithm}[t]
\caption{ADOBI Sampling}
\begin{algorithmic}[1]
\Require $\Rsf_\theta$, $\ybm$, Training Steps $N$, NFE steps $T$  
\State $\{\bar{\alpha}_i, \bar{\sigma}_i\}_{i=1}^T\gets \reschedule(\{\alpha_i, {\sigma}_t\}_{i=1}^N, T)$
\For{$t = T$ to $1$}
    \State $\xbm_{0|{t}} \gets \Rsf_\theta \left(\xbm_t, t \right)$
    \Statex $\text{\# Image update using~(\ref{eq:update_img})}$
    \State $\xbm_{0|{t}}' \gets \xbm_{0|t} - \gamma_1 \nabla_{\xbm_{0|{t}}} \| \ybm - \Pbm\Fbm\Sbm_{{t}}(\xbm_{0|{t}}) \|^2_2$
    \Statex $\text{\# Forward model update using~(\ref{eq:loss_csm})}$
    \State $\Sbm_{t-1} \gets \mathsf{CSMUpdate}(\Sbm_{t}) $
    \Statex $\text{\# Sampling step using~(\ref{eq:DDB})}$
    \State $\bm{\epsilon} \sim \mathcal{N}(0, \Ibm)$
    \State $\xbm_{t-1} \gets \mathsf{Resample} (\xbm_{0|{t}}, \xbm_t, \{\bar{\alpha}_i, \bar{\sigma}_{i}\}_{i=t-1}^{i=t})$
\EndFor
\State \textbf{return} $\xbm_{0|0}$
\end{algorithmic}
\label{alg:sampling}
\end{algorithm}

\section{Proposed Method}

\subsection{ADOBI Training}
\label{sec:initialization}

Prior work on DBs~\cite{indi2023, liu2023i2sb} has defined the diffusion process directly in the image domain, by mapping degraded images to the target clean images using~\eqref{eq:DDB}. Since measurements and images in general inverse problems are in different domains, one first needs to map the measurements to the image domain using a dedicated initialization strategy.
We consider two initialization strategies for ADOBI:
    (1) \textit{Zero-Filled Initialization}: In this approach, the starting distribution is initialized with zero-filled images ($\Abm^\Hsf\ybm$), a common practice in MRI, where missing k-space data is replaced with zeros. This initialization is computationally simple, providing a neutral starting point. 
    (2) \textit{GRAPPA Initialization}: To start with a higher-quality initialization, we consider using the output of the classical GRAPPA method~\cite{griswoldGeneralized2002}, which leverages auto-calibrating data from k-space to generate an approximate image. This initialization provides the diffusion process with a more accurate initial distribution, bridging the gap between the initial and target distributions. Our numerical results generally show that GRAPPA leads to improvements over the zero-filled initialization for DDB training.
    
Given the image domain initialization, we define the DB process $p(\xbm_t|\xbm_0, \zbm)$ as follows:
\begin{equation}
  \xbm_t = (1 - \alpha_t)\xbm_0 + \alpha_t \zbm  + \sigma_t \bm
  \epsilon, \quad \bm\epsilon \sim \Ncal(0, \Ibf),
  \label{eq:db_sde}
\end{equation}
where $\{\alpha_t, \sigma_t \}$ is the signal/noise schedule governing the evolution of the diffusion process over time $t$, $\zbm$ is the initialized image based on the k-space measurement $\ybm$. The network $\Rsf_{\thetabm}$ for ADOBI is trained to predict the posterior mean at each step by minimizing
\begin{equation}
\E_{(\xbm, \ybm) \sim p(\xbm, \ybm), t \sim p(t)}\left\| \Rsf_{\thetabm} \left(\xbm_t, t \right) - \xbm_0 \right\|^2_2.
\label{eq:db_training}
\end{equation}
The training pseudocode is provided in Algorithm~\ref{alg:training}.

\begin{figure*}[t]
    \centering
    \includegraphics[width=\linewidth]{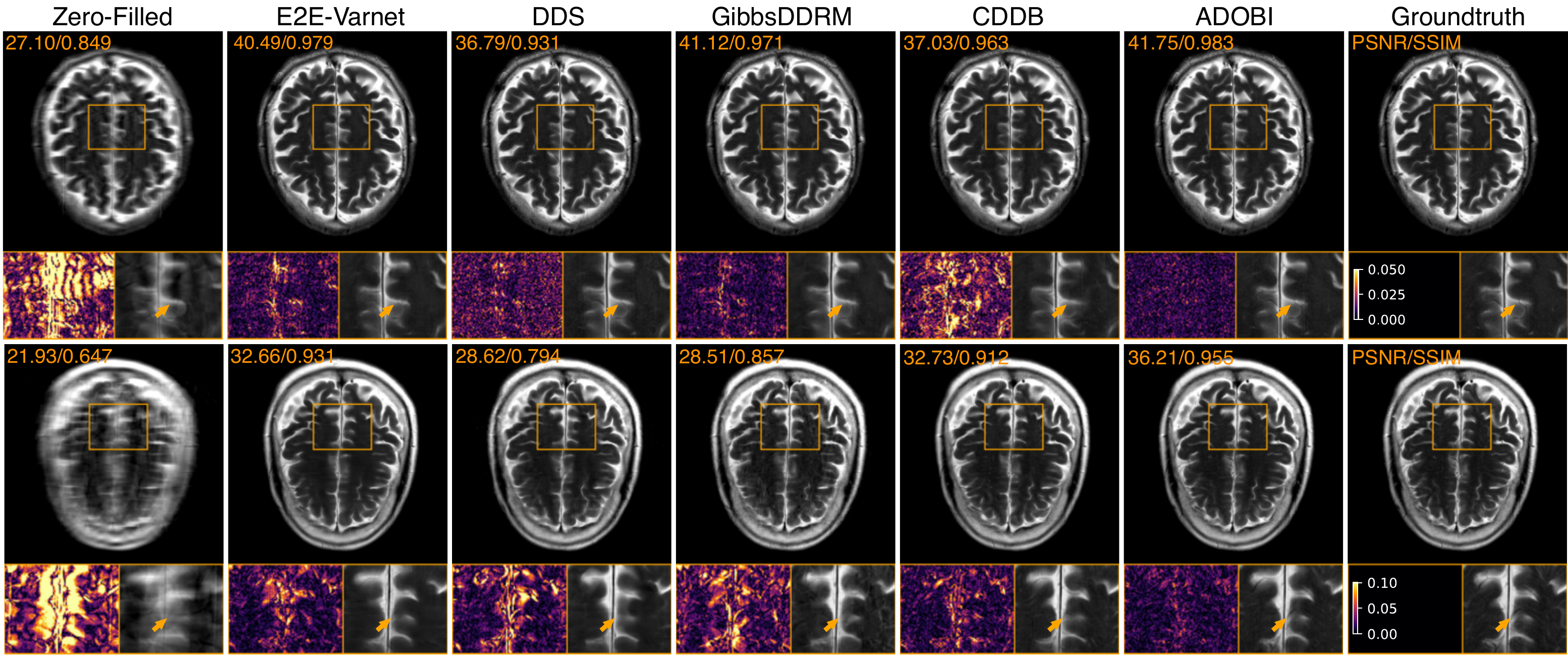}
    \caption{Visual illustration of ADOBI for MRI reconstruction compared to several baseline methods. The top row shows results for $4\times$ accelerated PMRI data collection while the bottom row shows those $8\times$. Error maps and zoomed in details highlight visual differences. Note how ADOBI provides the best visual and quantitative performance in both settings.}
\end{figure*}

\subsection{ADOBI Inference}
Previous DB-based methods, such as CDDB~\cite{chung2024direct}, assume complete knowledge of the forward measurement model. However, in parallel MRI (PMRI), the coil sensitivity maps of the forward model are often unknown. ADOBI addresses this limitation by introducing an adaptive mechanism that iteratively refines both image estimates and CSMs within a diffusion bridge framework, maintaining measurement consistency and significantly improving reconstruction fidelity.

As shown in Algorithm~\ref{alg:sampling}, ADOBI uses a two-step iterative update process for both the image and the measurement operator, leveraging diffusion bridge principles to enable efficient sampling and refinement. This adaptive inference ensures that the reconstructed image and the estimated measurement operator parameters (CSMs) remain consistent with the measurements throughout reconstruction.

\subsubsection{Image Update}
To enforce data consistency with the observed measurements 
$\ybm$, we apply an update rule directly to the posterior mean estimation $\xbm_0$ at each iteration to reduce the discrepancy with the observed data:
\begin{equation}
\xbm_{0|{t}}' \gets \xbm_{0|t} - \gamma_1 \nabla_{\xbm_{0|{t}}} \| \ybm - \Pbm\Fbm\Sbm_{{t}}(\xbm_{0|{t}}) \|^2_2,
\label{eq:update_img}
\end{equation}
where $\xbm_0$ represents the posterior mean estimated by the diffusion bridge network, and $\xbm'_{0|t}$ denotes the updated image that conforms to the forward model constraints. This iterative update reduces the error between the observed measurements $\ybm$, and the predicted measurements, enhancing the model’s fidelity to the original k-space data. 

However, as this update relies on the existing measurement model without directly addressing any of its inaccuracies, it necessitates a second, complementary update step.

\begin{table*}[!t]\small
\centering
\caption{Comparison of various methods for 4x and 8x acceleration in noise-free setting on fastMRI Brain images. ADOBI outperforms other baselines in both reconstruction accuracy and perceptual quality, with smaller standard deviation in the corresponding values. \colorbox{lightgreen}{\makebox(55,6){\textbf{Best values}}}~and~\colorbox{lightblue}{\makebox(80,6){second-best values}} for each metric are color-coded.}
\setlength{\tabcolsep}{2.9pt}
\begin{tabular}{c|ccc|ccc}
\toprule[1.2pt]
           & \multicolumn{3}{c}{4x Acceleration}   & \multicolumn{3}{c}{8x Acceleration}   \\ \midrule
Method    & PSNR($\uparrow$)   & SSIM($\uparrow$)    & LPIPS($\downarrow$)   & PSNR($\uparrow$)    & SSIM($\uparrow$)    & LPIPS($\downarrow$)   \\ \midrule
 Zero-filled       & $26.85\pm1.56$ & $0.844\pm0.033$ & $0.195\pm0.024$& $21.68\pm1.38$ & $0.644\pm0.057$ & $0.347\pm0.039$  \\
GRAPPA~\cite{griswoldGeneralized2002}       & $37.31\pm1.37$ & $0.967\pm0.007$ & $0.079\pm0.015$& $27.33\pm1.40$ & $0.801\pm0.048$ & $0.207\pm0.039$  \\
SwinIR~\cite{liang2021swinir}    & $36.11\pm1.38$ & $0.962\pm0.009$ & $0.075\pm0.014$ & $29.40\pm1.44$ & $0.898\pm0.019$ & $0.150\pm0.022$ \\
E2E-VarNet~\cite{sriramEndtoend2020} & $40.54\pm1.58$ & $0.982\pm0.005$ & $0.039\pm0.014$ & $32.45\pm1.49$ &$0.932\pm0.012$  & $0.130\pm0.022$ \\
DPS~\cite{chung2023diffusion}      & $33.74\pm1.36$ & $0.902\pm0.018$ & $0.102\pm0.030$ & $29.74\pm1.39$ & $0.834\pm0.028$ & $0.131\pm0.030$ \\
DDS~\cite{chungdecomposed}     & \colorbox{lightblue}{\makebox(\boxsize,4){$42.50\pm3.71$}} & \colorbox{lightblue}{\makebox(\boxsize,4){$0.979\pm0.045$}} & \colorbox{lightblue}{\makebox(\boxsize,4){$0.035\pm0.044$}} & $31.58\pm2.28$ & $0.893\pm0.051$ & $0.145\pm0.061$ \\ 
GibbsDDRM~\cite{murata2023gibbsddrm} &$42.44\pm4.67$ &$0.968\pm0.072$ &$0.037\pm0.052$ &$32.13\pm2.36$ &$0.895\pm0.059$ &$0.143\pm0.062$ \\


I2SB~\cite{liu2023i2sb}      & $33.29\pm0.99$ & $0.926\pm0.012$ & $0.127\pm0.018$ & $30.75\pm1.16$ & $0.896\pm0.018$ & $0.159\pm0.022$ \\
CDDB~\cite{chung2024direct}       & $38.89\pm1.30$ & $0.973\pm0.007$ & $0.045\pm0.010$ & $32.60\pm 1.35$ & $0.908\pm0.018$ & $0.140\pm0.019$ \\
\midrule  ADOBI (ZF)       & $40.29\pm1.43$ & $0.978\pm0.006$ & $0.039\pm0.010$ & \colorbox{lightblue}{\makebox(\boxsize,4){$33.51\pm1.59$}} & \colorbox{lightblue}{\makebox(60,4){$0.916\pm0.019$}} & \colorbox{lightblue}{\makebox(\boxsize,4){$0.125\pm0.0200$}} \\ 
ADOBI (G)       & \colorbox{lightgreen}{\makebox(\boxsize,4){$\mathbf{42.59\pm1.20}$}} & \colorbox{lightgreen}{\makebox(\boxsize,4){$\mathbf{0.984\pm0.004}$}} & \colorbox{lightgreen}{\makebox(\boxsize,4){$\mathbf{0.022\pm0.006}$}} & \colorbox{lightgreen}{\makebox(\boxsize,4){$\mathbf{36.18\pm1.58}$}} & \colorbox{lightgreen}{\makebox(\boxsize,4){$\mathbf{0.954\pm0.011}$}} & \colorbox{lightgreen}{\makebox(\boxsize,4){$\mathbf{0.054\pm0.013}$}} \\ \bottomrule[1.2pt]
\end{tabular}
\label{tb:main_noisefree}
\end{table*}

\begin{table*}[!t]\small
\centering
\caption{Comparison of various methods for 4x and 8x acceleration in noisy setting on fastMRI Brain images. ADOBI outperforms most of the other baselines in both reconstruction accuracy and perceptual quality, with smaller standard deviation in the corresponding values. \colorbox{lightgreen}{\makebox(\boxsize,6){\textbf{Best values}}}~and~\colorbox{lightblue}{\makebox(80,6){second-best values}} for each metric are color-coded.}
\setlength{\tabcolsep}{3pt}
\begin{tabular}{c|ccc|ccc}
\toprule[1.2pt]
           & \multicolumn{3}{c}{4x Acceleration}   & \multicolumn{3}{c}{8x Acceleration}   \\ \midrule
Method    & PSNR($\uparrow$)   & SSIM($\uparrow$)    & LPIPS($\downarrow$)   & PSNR($\uparrow$)    & SSIM($\uparrow$)    & LPIPS($\downarrow$)   \\ \midrule
 Zero-filled       & $26.23\pm1.53$ & $0.823\pm0.036$  &$0.204\pm0.024$ &$20.49\pm1.42$ & $0.618\pm0.059$ & $0.389\pm0.051$  \\
GRAPPA~\cite{griswoldGeneralized2002}       & $34.37\pm1.76$ & $0.883\pm0.044$ & $0.134\pm0.039$& $25.92\pm1.30$ & $0.674\pm0.067$ & $0.307\pm0.048$  \\
  
SwinIR~\cite{liang2021swinir} &$36.09\pm1.42$ &$0.960\pm0.010$ &$0.091\pm0.015$ &$29.20\pm1.24$ &$0.892\pm0.018$ &$0.183\pm0.024$ \\
E2E-VarNet~\cite{sriramEndtoend2020} &$39.16\pm1.74$ &$0.971\pm0.007$ &$0.046\pm0.015$ &$31.83\pm1.53$ &$0.929\pm0.013$ &$0.129\pm0.023$ \\
DPS~\cite{chung2023diffusion}        & $31.77\pm1.40$ & $0.876\pm0.019$ & $0.117\pm0.031$ & $28.52\pm1.22$ & $0.826\pm0.024$ & $0.146\pm0.028$ \\
DDS~\cite{chungdecomposed}     & $39.52\pm3.72$ & $0.950\pm0.064$ & \colorbox{lightgreen}{\makebox(\boxsize,4){$\mathbf{0.034\pm0.048}$}} & $31.37\pm2.40$ & $0.883\pm0.056$ & $0.131\pm0.046$ \\


GibbsDDRM~\cite{murata2023gibbsddrm}       & \colorbox{lightblue}{\makebox(\boxsize,4){$39.70\pm3.48$}} & $0.951\pm0.081$ & $0.052\pm0.060$ & $31.90\pm2.34$ & $0.891\pm0.074$ & $0.108\pm0.054$  \\
I2SB~\cite{liu2023i2sb}      & $32.96\pm1.05$ & $0.923\pm0.013$ & $0.094\pm0.018$ & $30.74\pm1.14$ & $0.899\pm0.018$ & $0.121\pm0.021$ \\
CDDB~\cite{chung2024direct}       & $36.35\pm1.16$ & $0.943\pm0.014$ & $0.079\pm0.015$ & $31.33\pm1.24$ & $0.835\pm0.025$ & $0.161\pm0.025$ \\

\midrule
ADOBI (ZF)       &  $39.58\pm1.34$ & \colorbox{lightblue}{\makebox(\boxsize,4){$0.975\pm0.006$}} & $0.044\pm
0.010$  & \colorbox{lightblue}{\makebox(\boxsize,4){$33.39\pm1.50$}} & \colorbox{lightblue}{\makebox(\boxsize,4){$0.916\pm0.019$}} & \colorbox{lightblue}{\makebox(\boxsize,4){$0.126\pm0.020$}} \\
ADOBI (G)      & \colorbox{lightgreen}{\makebox(\boxsize,4){$\mathbf{40.89\pm1.07}$}} & \colorbox{lightgreen}{\makebox(\boxsize,4){$\mathbf{0.979\pm0.005}$}} & \colorbox{lightblue}{\makebox(\boxsize,4){$0.039\pm0.009$}} & \colorbox{lightgreen}{\makebox(\boxsize,4){$\mathbf{34.28\pm1.34}$}} & \colorbox{lightgreen}{\makebox(\boxsize,4){$\mathbf{0.942\pm0.011}$}} & \colorbox{lightgreen}{\makebox(\boxsize,4){$\mathbf{0.085\pm0.015}$}} \\ \bottomrule[1.2pt]
\end{tabular}
\label{tb:main_noisy}
\end{table*}

\subsubsection{Forward Model Update}
\label{sec:forward_update}

In addition to refining the image estimate, ADOBI updates the coil sensitivity maps $\Sbm$ using the image samples $\xbm_0$, adapting the forward model based on current reconstruction. This update is formulated as:
\begin{equation}
\Sbm_{t} = \argmin_{\Sbm_t}\| \ybm - \Pbm\Fbm\Sbm_{t}(\xbm_{0|{t}}) \|^2_2 + \lambda ||\Sbm_{t} - \Sbm_\text{inital}||_2^2,
\label{eq:loss_csm}
\end{equation}
where $\Sbm_\text{inital}$ denotes the ESPIRiT estimation of $\Sbm$, which incorporates a Tikhonov penalty as the prior for CSM estimates. By jointly optimizing the image and CSMs, ADOBI dynamically adapts to uncertainties in the forward model, improving data alignment and minimizing artifacts. A gradient descent algorithm iteratively minimizes this objective, allowing the forward model to adjust to image updates and reinforcing the data fidelity and regularization requirements. This joint optimization ensures a robust solution even under partially known measurement models, enhancing overall image quality and fidelity to the measurement. Our numerical results show that the measurement operator update has a negligible effect on on the overall computational complexity of the reconstruction (see \autoref{tb:ablation_calibration}).

\section{Experimental Validation}

\subsection{Dataset}
We consider multicoil, T2-weighted human brain images from the fastMRI~\cite{zbontarFastMRI2018} dataset, selecting 4,912 images for training and 470 for testing. To ensure image quality, we excluded the first four and last five slices of each volume. Each complex-valued image is then cropped to a consistent shape of $320\times320$. We simulated Cartesian under-sampling masks with acceleration rates of $4\times$, and $8\times$.  The coil sensitivity maps (CSMs) $\Sbm$ are estimated using the ESPIRiT~\cite{ueckerESPIRiTan2014} algorithm: for ground truth, $\Sbm$ is obtained from fully sampled data, while initial CSMs for each undersampled setting are estimated from the corresponding undersampled data. The GRAPPA warm-up reconstruction is generated using the code in this link\footnote{https://github.com/mckib2/pygrappa}. For noisy settings, we add Gaussian noise to the undersampled measurement whose norm is $10\%$ of of the measurement data.


\subsection{Baselines}
We compare ADOBI with various diffusion model (DM)-based and diffusion bridge (DB)-based methods for MRI reconstruction tasks. Among DB methods, we include comparisons with DPS~\cite{chung2023diffusion} and DDS~\cite{chungdecomposed}, both of which utilize pre-trained diffusion models as priors, but employing distinct posterior sampling strategies. For DB methods, we benchmark our proposed method against I2SB~\cite{liu2023i2sb} and CDDB~\cite{chung2024direct}, which represent diffusion bridges without and with measurement constraints, respectively. Note that, I2SB and CDDB share the same DDB backbone used in ADOBI. We also compare our method with GibbsDDRM~\cite{murata2023gibbsddrm}, a method that integrates diffusion models with Gibbs sampling for image reconstruction and simultaneous forward model estimation in blind inverse problems. Additionally, we evaluate two widely-used reconstruction methods, SwinIR~\cite{liang2021swinir} and E2E-VarNet~\cite{Sriram.etal2020}, to further highlight the effectiveness of ADOBI. Note that E2E-VarNet~\cite{Sriram.etal2020} is an MDBL network that includes a separate model for automatic CSM estimation. Note that in all the baseline methods, CSMs are initialized using ESPIRiT pre-estimated CSMs for consistency across comparisons. Additional implementation details of baseline method are provided in the appendix.

We evaluated ADOBI against all baseline methods under two acceleration rates ($4\times$ and $8\times$) in both noise-free and noisy settings. As shown in \autoref{tb:main_noisefree} and \autoref{tb:main_noisy}, ADOBI consistently outperforms all baseline methods across both initialization types (zero-filled and GRAPPA warm-up), demonstrating the effectiveness of our joint sampling updates for both the image and CSMs in the forward operator. Importantly, ADOBI achieves these results with remarkable speed, requiring only 10 NFEs to surpass state-of-the-art DM-based methods DDS~\cite{chungdecomposed} in both perception and distortion performance that need over 100 NFEs, showcasing a significant advantage in efficiency.

\begin{figure}[ht]
    \centering    \includegraphics[width=3.06in]{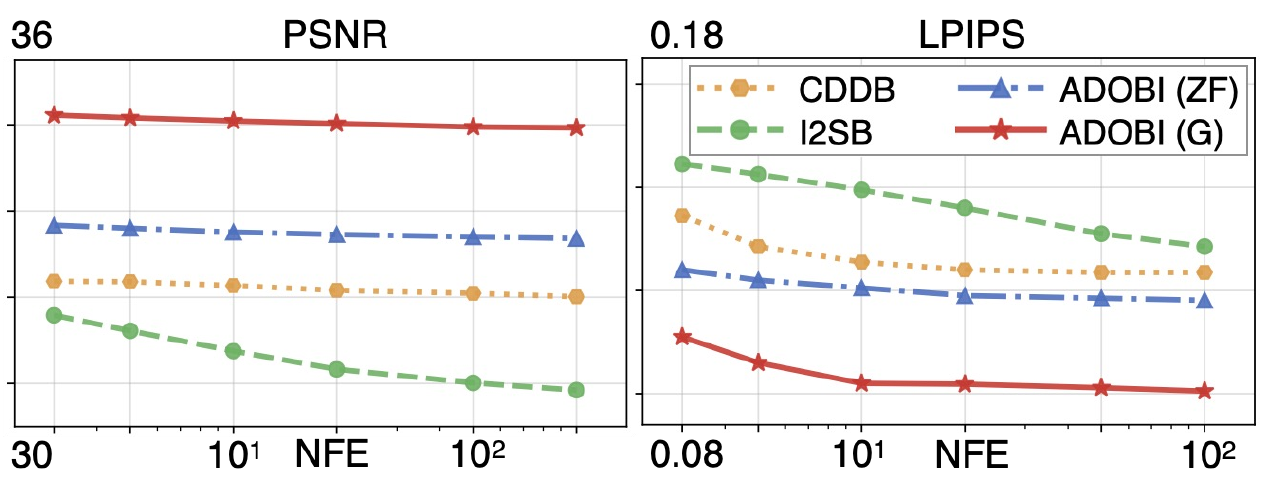}
    \caption{Perception and distortion performance comparison at $8\times$ acceleration across different NFEs. Left: NFE vs. PSNR; Right: NFE vs. LPIPS. Our method outperforms baseline methods across most NFE settings, achieving superior perception and distortion performance.}
    \label{fig:ablation_nfe}
\end{figure}

Another ADOBI’s key strength is its ability to balance perceptual quality with low distortion, outperforming other DB-based methods on this Pareto frontier.  As shown in \autoref{fig:ablation_nfe}, ADOBI provides a more favorable perception-distortion trade-off, and consistently achieving superior performance in both distortion-based and perception-based metrics. This balance demonstrates ADOBI’s versatility in offering high-quality reconstructions adaptable to applications where visual clarity is as important as accuracy.

\begin{table}[]\small
\centering
\setlength{\tabcolsep}{4pt}
\caption{Ablation studies on initialization strategies for DDB backbone training in $4\times$ and $8\times$ PMRI reconstruction. Our adaptive training approach leveraging GRAPPA initialization enhances performance for both ADOBI and CDDB methods. \colorbox{lightgreen}{\makebox(55,6){\textbf{Best values}}}~and~\colorbox{lightblue}{\makebox(78,6){second-best values}} for each metric are colored.}
\begin{tabular}{ccccc}
\toprule[1.2pt]
Acce  & Initialization & PSNR($\uparrow$)   & SSIM($\uparrow$)    & LPIPS($\downarrow$)      \\
\midrule
$4\times$   &$\text{CDDB}_\text{Zero-filled}$   & $38.89$ & $0.973$ & $0.045$ \\
$4\times$   &$\text{ADOBI}_\text{Zero-filled}$ & \colorbox{lightblue}{\makebox(20,6){$40.29$}} & \colorbox{lightblue}{\makebox(20,6){$0.978$}} & \colorbox{lightblue}{\makebox(20,6){$0.039$}} \\
$4\times$   & $\text{CDDB}_\text{GRAPPA}$   & $39.63$ & $0.974$ & $0.059$ \\
$4\times$   & $\text{ADOBI}_\text{GRAPPA}$   & \colorbox{lightgreen}{\makebox(20,6){$\mathbf{42.59}$}} & \colorbox{lightgreen}{\makebox(20,6){$\mathbf{0.984}$}} & \colorbox{lightgreen}{\makebox(20,6){$\mathbf{0.022}$}} \\

\midrule
$8\times$   &$\text{CDDB}_\text{Zero-filled}$   & $32.60$ & $0.908$ & $0.140$ \\
 $8\times$   &$\text{ADOBI}_\text{Zero-filled}$ & $33.51$ & $0.916$ & $0.131$ \\
$8\times$   & $\text{CDDB}_\text{GRAPPA}$   & \colorbox{lightblue}{\makebox(20,6){$35.38$}} & \colorbox{lightblue}{\makebox(20,6){$0.934$}} & \colorbox{lightblue}{\makebox(20,6){$0.109$}} \\
$8\times$   & $\text{ADOBI}_\text{GRAPPA}$   & \colorbox{lightgreen}{\makebox(20,6){$\mathbf{36.18}$}} & \colorbox{lightgreen}{\makebox(20,6){$\mathbf{0.954}$}} & \colorbox{lightgreen}{\makebox(20,6){$\mathbf{0.080}$}} \\
\bottomrule[1.2pt]
\end{tabular}
\label{tb:ablation_warmup_8x}
\end{table}

\subsection{Ablation Study}
To rigorously assess the contributions of each component in this work, we performed several ablation studies. Due to the space limit, we moved several more ablation studies to the appendix.

\subsubsection{Effectiveness of Forward Model Calibration}
This section explores the impact of forward model calibration step on reconstruction quality of ADOBI (See Section~\ref{sec:forward_update}). We evaluated three ADOBI variants: $\text{ADOBI}_\text{w/o Calibration}$, $\text{ADOBI}_\text{w/o Calibration}$, and $\text{ADOBI}_\text{Groundtruth CSMs}$. Specifically, $\text{ADOBI}_\text{w/o Calibration}$ and $\text{ADOBI}_\text{Groundtruth CSMs}$ use pre-estimated and ground-truth measurement operators, respectively. \autoref{tb:ablation_calibration} shows that ADOBI with a calibrated forward model consistently outperforms the non-calibrated variant and nearly matches the performance of ADOBI with Groundtruth CSMs. Additionally, the runtime difference between calibrated and non-calibrated ADOBI is minimal, indicating that forward model updates have negligible impact on inference speed. These results underscore the importance of forward model calibration in ADOBI.

\subsubsection{Influence of initializations for ADOBI backbone} In this subsection, we examine how different initialization strategies influence the performance of the ADOBI's DB backbone (See Section~\ref{sec:initialization}). As shown in \autoref{tb:ablation_warmup_8x}, both the CDDB and our proposed sampling strategy demonstrate improved performance when using GRAPPA initialization, which provides a closer approximation to the ground-truth distribution than zero-filled initialization. This improvement arises because GRAPPA performs pre-interpolation in k-space, effectively reducing ghosting artifacts in the image domain caused by missing k-space data. This demonstrates our proposed training strategy use the flexibility of DB in building the generative process from a range of initial distributions, surpassing the traditional Gaussian noise-based initialization for diffusion models.

\begin{table}[ht]\small
\centering
\caption{Ablation study to verify the effectiveness of forward model calibration in ADOBI. Note how the use of forward model calibration over the CSMs enables $\text{ADOBI}_\text{w/ Calibration}$ to outperform
$\text{ADOBI}_\text{w/o Calibration}$ and approach the performance of the oracle algorithm $\text{ADOBI}_\text{w/ Groundtruth CSMs}$. Here, runtime indicates inference time in seconds per image for each method, with NFE=10.}
\setlength{\tabcolsep}{3pt}
\begin{tabular}{cccccc}
\toprule[1.2pt]
Acce.  & Method & PSNR   & SSIM    & LPIPS & Runtime      \\ \midrule
$4\times$   &$\text{ADOBI}_\text{w/o Calibration}$ & $40.72$ & $0.974$ & $0.043$  &$1.92$ s\\
$4\times$   &$\text{ADOBI}_\text{w/ Calibration}$ & $40.89$ & $0.979$ & $0.039$ &$2.62$ s \\
$4\times$   &$\text{ADOBI}_\text{w/ Groundtruth CSMs}$ & $41.03$ & $0.980$ & $0.037$ &$1.92$ s \\
\midrule
$8\times$   &$\text{ADOBI}_\text{w/o Calibration}$ & $33.92$ & $0.929$ & $0.111$ &$1.92$ s\\
$8\times$   &$\text{ADOBI}_\text{w/ Calibration}$ & $34.28$ & $0.942$ & $0.085$ &$2.62$ s\\
$8\times$   &$\text{ADOBI}_\text{w/ Groundtruth CSMs}$ & $34.34$ & $0.942$ & $0.083$ &$1.92$ s\\
\bottomrule[1.2pt]
\end{tabular}
\label{tb:ablation_calibration}
\end{table}


\subsubsection{Influence of Stochastic Perturbation}
We explore the role of stochastic perturbation within the ADOBI framework for PMRI reconstruction. In our formulation, disabling stochastic perturbation transforms the approach from an SDE-based to an ODE-based framework. Specifically, this is achieved by setting $\sigma_t$ in the forward process of the Diffusion Bridge (as shown in Equation~\ref{eq:db_sde}) to zero. As shown in \autoref{tb:ablation_ode}, the SDE-based approach performs better than the ODE-based approach for PMRI reconstruction. Our experimental findings indicate that the SDE-based approach consistently outperforms the ODE-based alternative for PMRI reconstruction.

\section{Discussion}
In this section, we further discuss the benefits and insights provided by ADOBI.

\subsection{Stability of ADOBI Relative to DM Methods}
\begin{figure}
    \centering
    \includegraphics[width=3.2in]{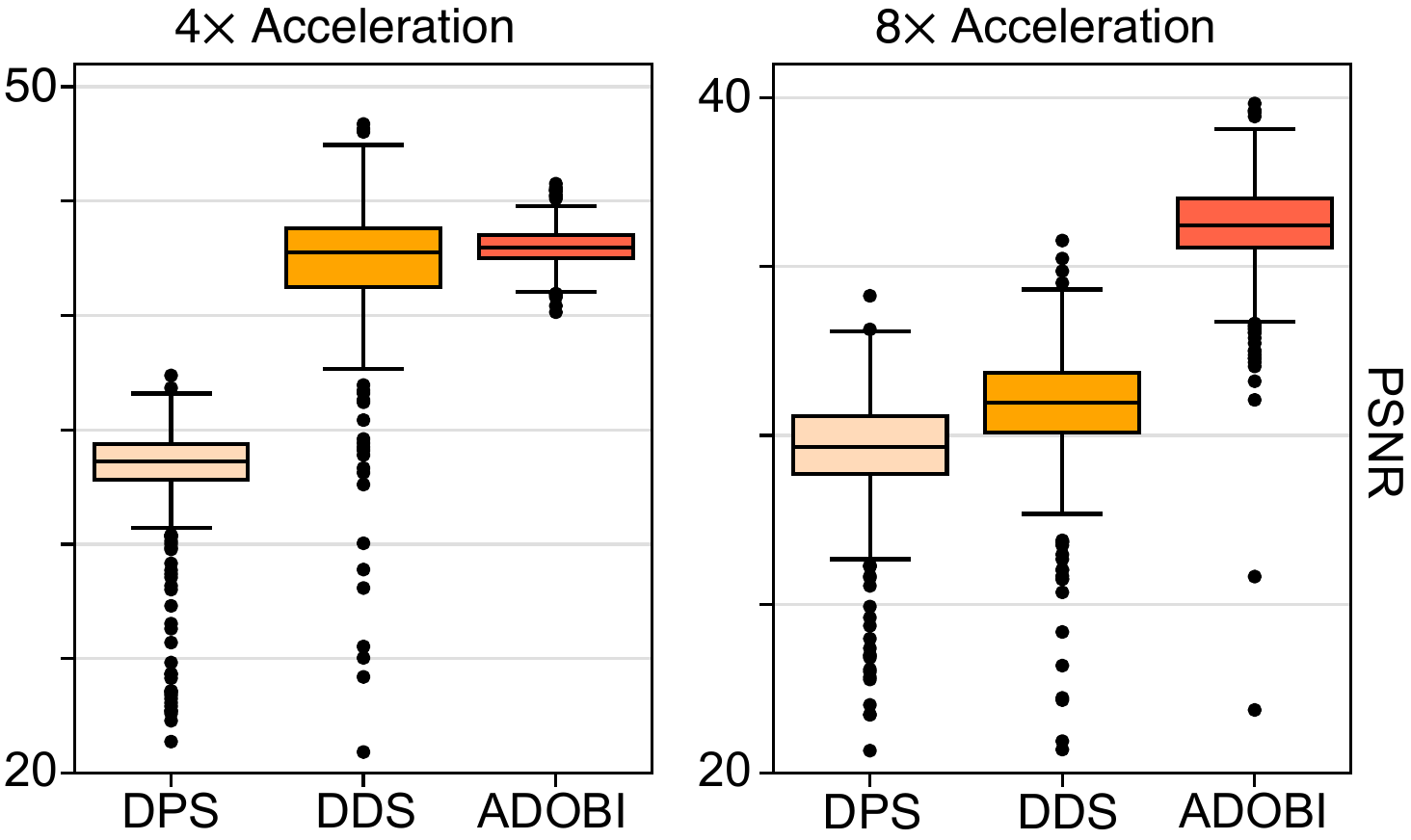}
    \caption{Box plot comparison of baseline methods at 4x acceleration (left) and 8x acceleration (right). Whiskers show upper and lower bounds, with dots as outliers. Notably, ADOBI shows fewer outliers, highlighting its stability.}
    \label{fig:box_plot}
\end{figure}

\begin{figure*}[h]
    \centering
    \includegraphics[width=5.88in]{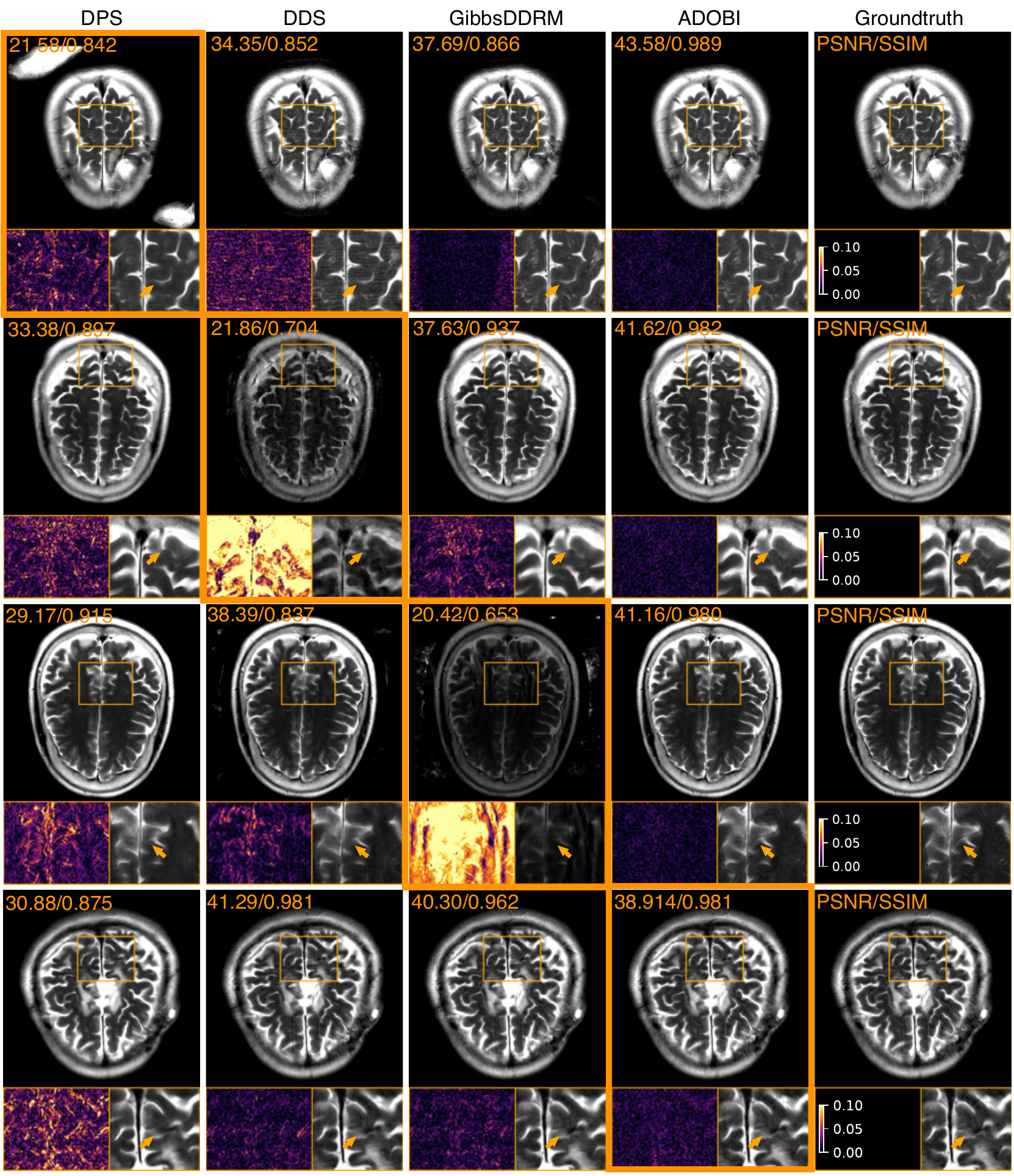}
    \caption{Worst case comparison on 4x acceleration MRI images. The rows from top to bottom denote the lowest-PSNR slices for DPS, DDS, GibbsDDRM, and ADOBI, respectively. We box the particular worst-case result in orange, and display corresponding results from other methods on the same slice for each row.}
        \label{fig:worst_case}
\end{figure*}

We assess the stability of ADOBI in comparison to traditional diffusion model-based methods. We assess the worst-case performance of ADOBI compared to the state-of-the-art DM baselines, DPS and DDS. As shown in \autoref{fig:worst_case}, while DDS and DPS exhibit significant artifacts and degraded image quality in their worst-case outputs, ADOBI consistently preserves essential details under the same challenging conditions, which is curial for medical imaging application. \autoref{fig:box_plot} further highlights ADOBI's robust and superior average performance, exhibiting fewer outliers compared to both DDS and DPS.  This robustness makes ADOBI a reliable alternative for applications demanding precise and stable reconstructions.

\begin{table}[]\small
\centering
\caption{Ablation study on the effectiveness of the ODE and SDE DB backbones in ADOBI for PMRI. The SDE-based DB consistently leads to better performance in both $4\times$ and $8\times$ settings.}
\setlength{\tabcolsep}{5pt}
\begin{tabular}{ccccc}
\toprule[1.2pt]
Acce.  & Type & PSNR($\uparrow$)   & SSIM($\uparrow$)    & LPIPS($\downarrow$)      \\ \midrule
$4\times$   &$\text{ADOBI}_\text{ODE}$ & $39.42$ & $0.971$ & $0.051$ \\
$4\times$   &$\text{ADOBI}_\text{SDE}$  & $40.89$ & $0.979$ & $0.039$ \\
$8\times$   &$\text{ADOBI}_\text{ODE}$ & $33.11$ & $0.931$ & $0.092$ \\
$8\times$   &$\text{ADOBI}_\text{SDE}$ & $34.28$ & $0.942$ & $0.085$ \\
\bottomrule[1.2pt]
\end{tabular}
\label{tb:ablation_ode}
\end{table}

\begin{figure}[t]
    \centering
    \includegraphics[width=3.5in]{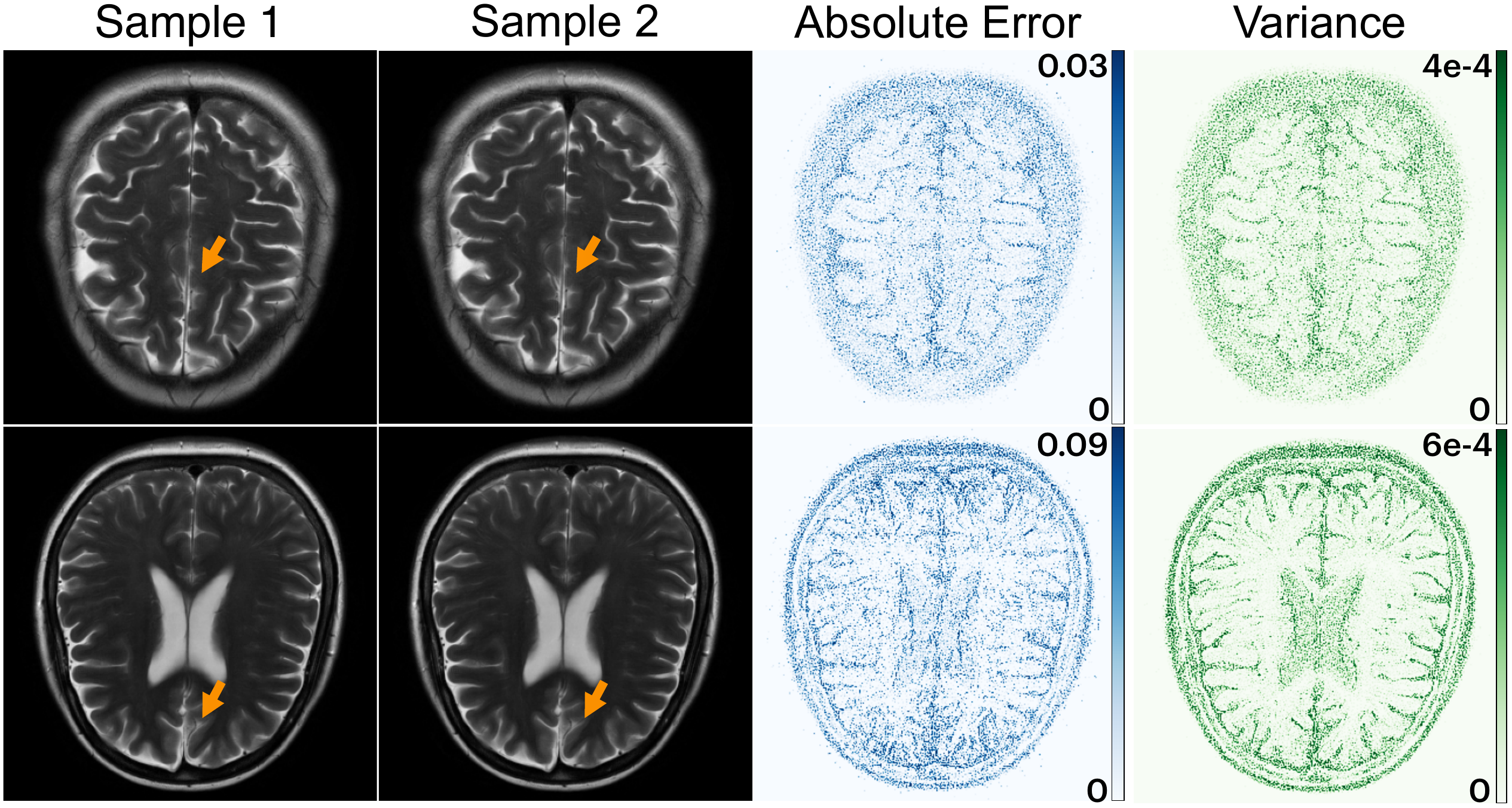}
    \small\caption{Uncertainty quantification results on 4x (top row) and 8x (bottom row) acceleration MRI images. Absolute error to ground truth corresponds to conditional mean $\mathbb{E}[\bm{x}|\bm{y}]$, and the variance is calculated by pixel-wise standard deviation. Ill-posed nature of the task has a direct effect on the diversity of generated images, and variance is heavily correlated to reconstruction errors.}
        \label{fig:uncertainty}
\end{figure}

\subsection{Uncertainty Quantification}
\autoref{fig:uncertainty} demonstrates ADOBI’s capability to quantify uncertainty by directly estimating variance. In a well-calibrated model, regions with larger absolute error correspond to areas of higher variance, allowing variance to serve as a proxy for reconstruction error when ground truth data is unavailable. As shown in \autoref{fig:uncertainty}, there is a strong correlation between the variance images and absolute error images, with higher uncertainty observed in regions with larger errors. Additionally, our method adapts to the ill-posed nature of the reconstruction task, as more challenging input conditions lead to increased variance in the output samples, reflecting greater uncertainty in those regions.

\section{Conclusion}
In this work, we introduced ADOBI, an adaptive diffusion bridge framework for blind inverse problems that, for the first time, can enforce measurement consistency but without the knowledge of forward model. ADOBI employs a adaptive process to jointly optimize both the image and forward model estimates, significantly enhancing reconstruction fidelity and reducing artifacts arising from forward model inaccuracies. Our results demonstrate that ADOBI outperforms existing PMRI reconstruction methods, achieving high-quality, robust reconstructions in only 5-10 steps.

\section{Acknowledgment}
This paper is supported by the NSF CAREER awards under grant CCF-2043134.

{\small
	\bibliographystyle{IEEEtran}
	\bibliography{main}
}

\newpage
\section{Appendix}

\subsection{Implementation details for ADOBI}

We train our model with PyTorch using code modified from I2SB\footnote{https://github.com/NVlabs/I2SB}~\cite{liu2023i2sb}. To UNet architecture setting we use the same setting as DDS~\cite{chungdecomposed} used for MRI setting. To accommodate the complex-valued PMRI images, we split the real and imaginary components into two channels, modifying the network to support a two-channel input and two-channel output format. We maintain 1000 training time steps and use the Adam optimizer with learning rate $5\times 10^{-5}$ and gamma weight decay of 0.99 every 1000 steps. We trained separate models for different initializations, accelerations, and noise settings, each with batch size of 2 and 400,000 iterations on a single NVIDIA A6000 GPU. 

\begin{figure}[h]
\centering
\includegraphics[width=2.5in]
{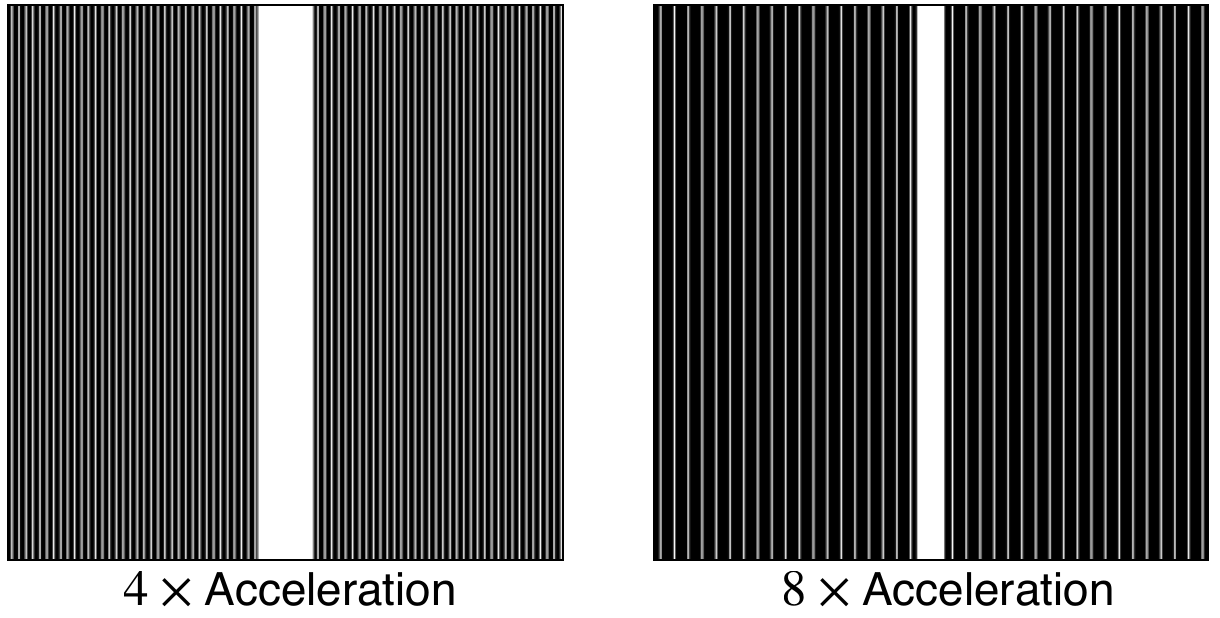}

\caption{Illustration of the undersampling masks used for CS-MRI in this work.}
\end{figure}

\subsection{Implementation details for Baselines}
For a fair comparison, we re-trained the diffusion model backbone for all the diffusion model based baselines with the same training data and model architecture setting we use for our proposed fine-turn. For DPS, we set the NFEs to 1000 as it suggested and fineturn the step-size for measurement constrain to the best. For DDS, we set the NFEs to 150 as it suggested and fine-turn the parameter for its Conjugate Gradient step to the best. For I2SB~\cite{liu2023i2sb} and CDDB~\cite{chung2024direct} they share the same diffusion bridge backbone with our proposed method. For SwinIR~\cite{liang2021swinir} and E2E-VarNet~\cite{sriramEndtoend2020}, we adopted the default architecture setting as they suggested in their paper.

\subsection{Inference Speed}
As shown in Table~\ref{tb:inference_speed}, we report the inference speed of each baseline method alongside our proposed ADOBI. Notably, ADOBI achieves state-of-the-art performance with only a minimal trade-off in speed compared to the other two diffusion bridge methods, I2SB and CDDB. Additionally, ADOBI demonstrates significantly faster inference compared to diffusion model-based methods.

\begin{table}[h]
\centering
\caption{The inference speed of each baseline method, measured in seconds per slice reconstruction.}
\begin{tabular}{ccr}
\toprule[1.2pt]
 Method  & NFE & Runtime      \\ \midrule
SwinIR  & 1 &$0.03$s\\
E2E-Varnet  & 1 &$0.16$s \\
DPS  &$1000$ &$194.60$s \\
DDS &$100$ &$16.86$s \\
GibbsDDRM  &$100$ &$19.45$s \\
I2SB &$10$ &$1.36$s \\
CDDB  &$10$ &$1.92$s \\ \midrule
ADOBI &$10$ &$2.62$s\\
\bottomrule[1.2pt]
\end{tabular}
\label{tb:inference_speed}
\end{table}

\subsection{Influence of Stochastic Perturbation for PMRI}
As discussed in the main paper, we evaluate the effect of stochastic perturbation on ADOBI for PMRI task.

In our formulation, disabling stochastic perturbation transforms the approach from an SDE-based to an ODE-based framework. Specifically, this is achieved by setting $\sigma_t$ in the forward process of the Diffusion Bridge (as shown in Equation~\ref{eq
}) to zero.  Under this ODE-based setting, the DB network is also trained to perform MMSE estimation, $\E[\xbm_{0}|\xbm_{t}]$, of the target image, following the same training approach as in the SDE setting (see Equation~\ref{eq:db_training}). In contrast to the SDE-based approach, as shown in Equation~\ref{eq
}, the ODE-based inference follows a deterministic process, expressed as: \begin{equation} \label{eq
} \begin{matrix} \E[\xbm_{t-1}|\xbm_{t}] = \left(1 - \frac{\alpha_{t-1}}{\alpha_t}\right)\E[\xbm_{0}|\xbm_{t}] + \left(\frac{\alpha_{t-1}}{\alpha_t}\right) \xbm_t. \end{matrix} \end{equation}

Our experimental findings suggest that the SDE-based approach performs better than the ODE-based approach for PMRI reconstruction. Our experimental findings indicate that the SDE-based approach consistently outperforms the ODE-based alternative for PMRI reconstruction. The potential reason is the SDE-based formulation can better capture the inherent distributional characteristics of MRI data by simulating a gradual transition between noisy and clean states. This gradual denoising process, supported by stochasticity, enables the model to retain and refine fine structural details that might otherwise be lost in a purely deterministic ODE approach.

\subsection{Comparison with Conditional Diffusion Model}

\begin{figure}[h]
\centering
    \includegraphics[width=3.5in]{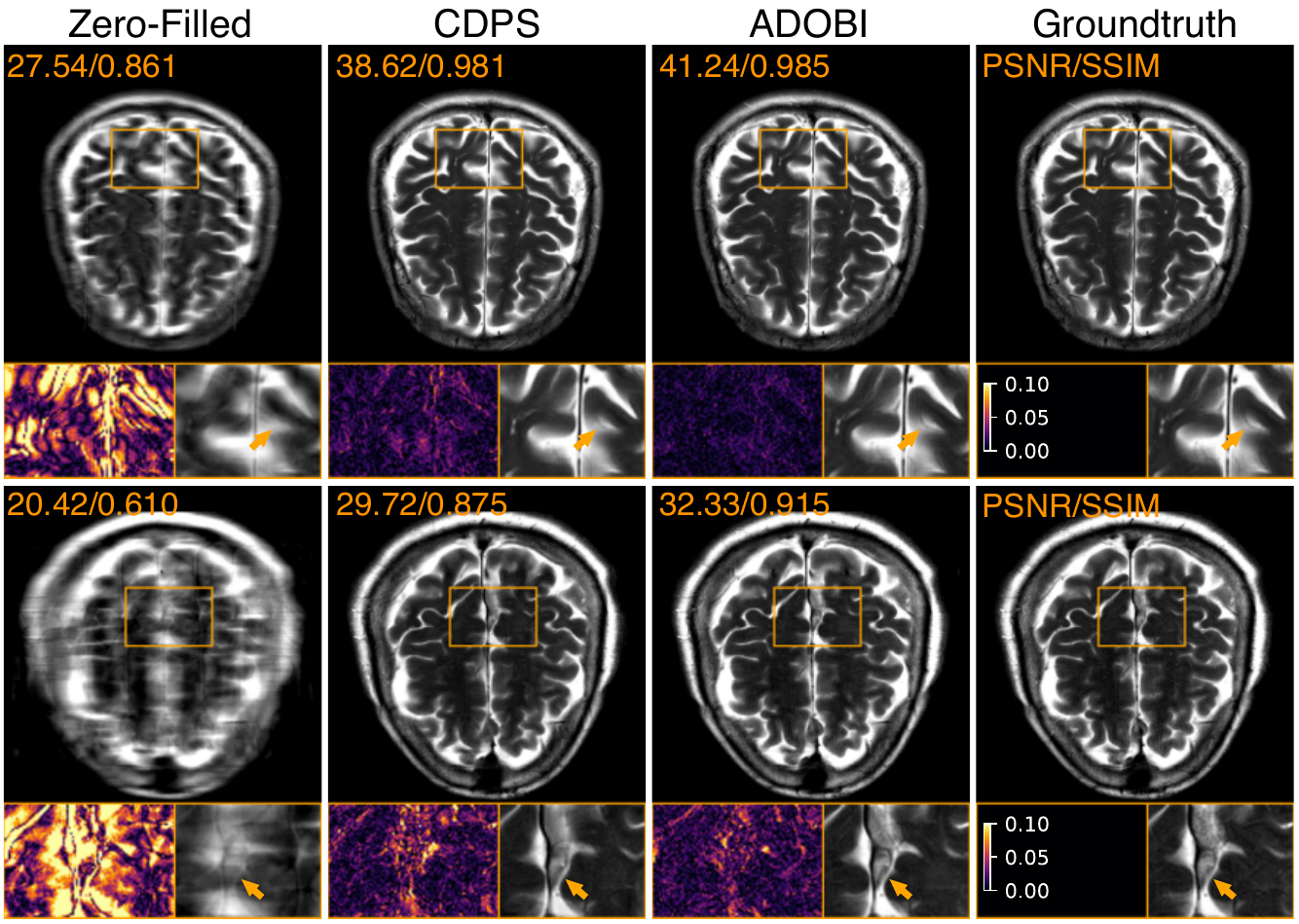}
\caption{Visual Comparison of ADOBI and conditional DPS (CDPS). The top row shows results for $4\times$ PMRI reconstruction while the bottom row shows $8\times$. Error maps and zoomed in details highlight visual differences. Note how ADOBI provides the best visual and quantitative performance in both settings.}
\label{fig:cdps}
\end{figure}

We further compared ADOBI with a conditional diffusion model-based method, using the same zero-filled warm-up as conditional inputs and applying DPS~\cite{chung2023diffusion} for sampling (referred to here as CDPS). As shown in Table~\ref{tb:cond_diffusion}, our diffusion bridge-based sampling outperforms the conditional diffusion model approach, demonstrating that the efficiency of ADOBI comes from our carefully designed framework rather than the warm-up process alone. Furthermore, Figure~\ref{fig:cdps} illustrates that ADOBI delivers superior visual performance, preserving finer details of brain tissue. Notably, ADOBI's inference speed is approximately $75\times$ faster than CDPS. These findings highlights the potential of diffusion bridge-based methods for advancing MRI reconstruction.

\begin{table}[t]\small
\centering
\caption{Comparison of ADOBI under zero-filled (ZF) and GRAPPA (G) initialization with conditional diffusion models for $4\times$ and $8\times$ acceleration in noise-free setting on fastMRI Brain images.}
\setlength{\tabcolsep}{4pt}
\begin{tabular}{ccccc}
\toprule[1.2pt]
Acce.  & Method & PSNR   & SSIM    & LPIPS      \\ \midrule
$4\times$   &Conditional DPS &  $39.67$ &$0.984$ &$0.039$ \\
$4\times$   &ADOBI (ZF) & $40.29$ & $0.978$ & $0.039$  \\
\midrule
$8\times$   &Conditional DPS & $32.91$ &$0.904$ &$0.129$ \\
$8\times$   &ADOBI (ZF) & $33.51$ & $0.916$ & $0.125$ \\
\bottomrule[1.2pt]
\end{tabular}
\label{tb:cond_diffusion}
\end{table}

\subsection{Ablation Study on Hyper-parameter}

\begin{figure}[h]
\centering    \includegraphics[width=2.5in]{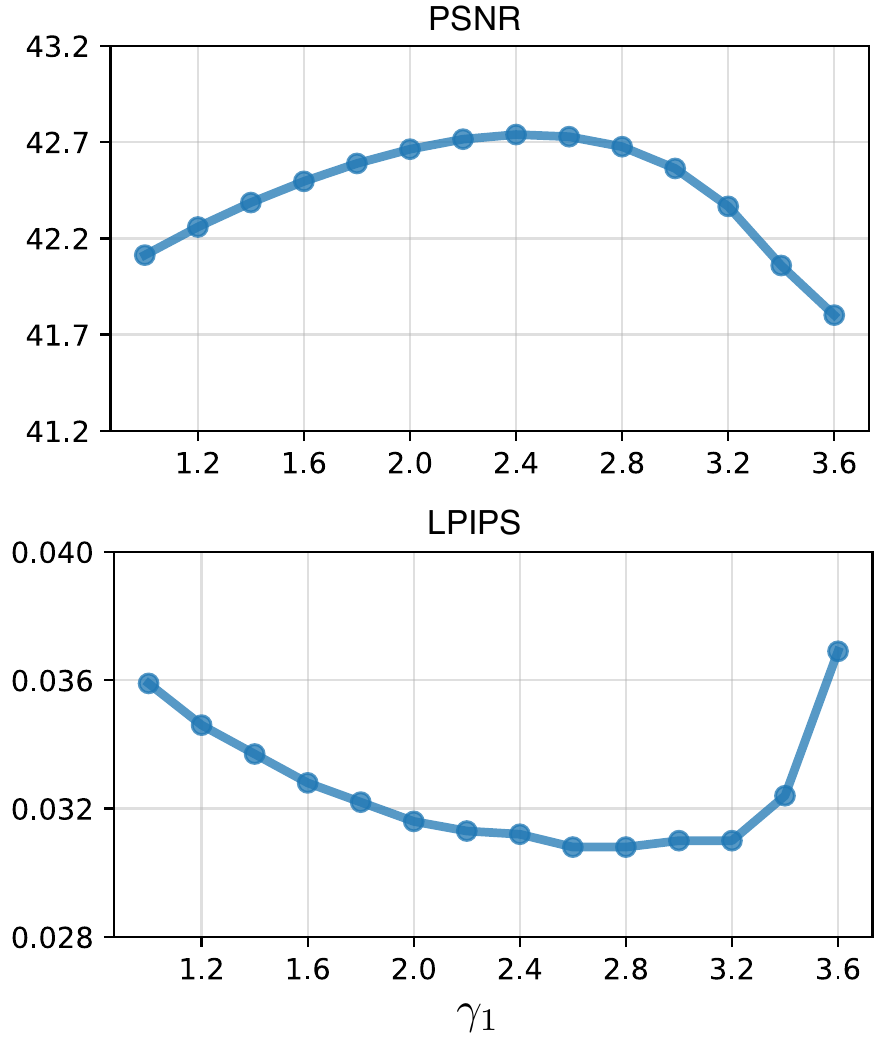}
\caption{ADOBI's performance comparison at $4\times$ acceleration setting with varying $\gamma$. Top: PSNR vs. $\gamma$; Bottom: LPIPS vs. $\gamma$.}
\label{fig:gamma_psnr}
\end{figure}

To evaluate the impact of the step-size hyperparameter $\gamma$, we conducted an ablation study on the $4\times$ CS-MRI task. As illustrated in Figure~\ref{fig:gamma_psnr}, the performance of ADOBI varies with different values of $\gamma$. In this experiment, $\gamma = 2.4$ yielded the best results in terms of both PSNR and LPIPS, highlighting its importance in achieving optimal reconstruction performance.

\begin{figure*}[h]
    \centering
    \includegraphics[width=5.88in]{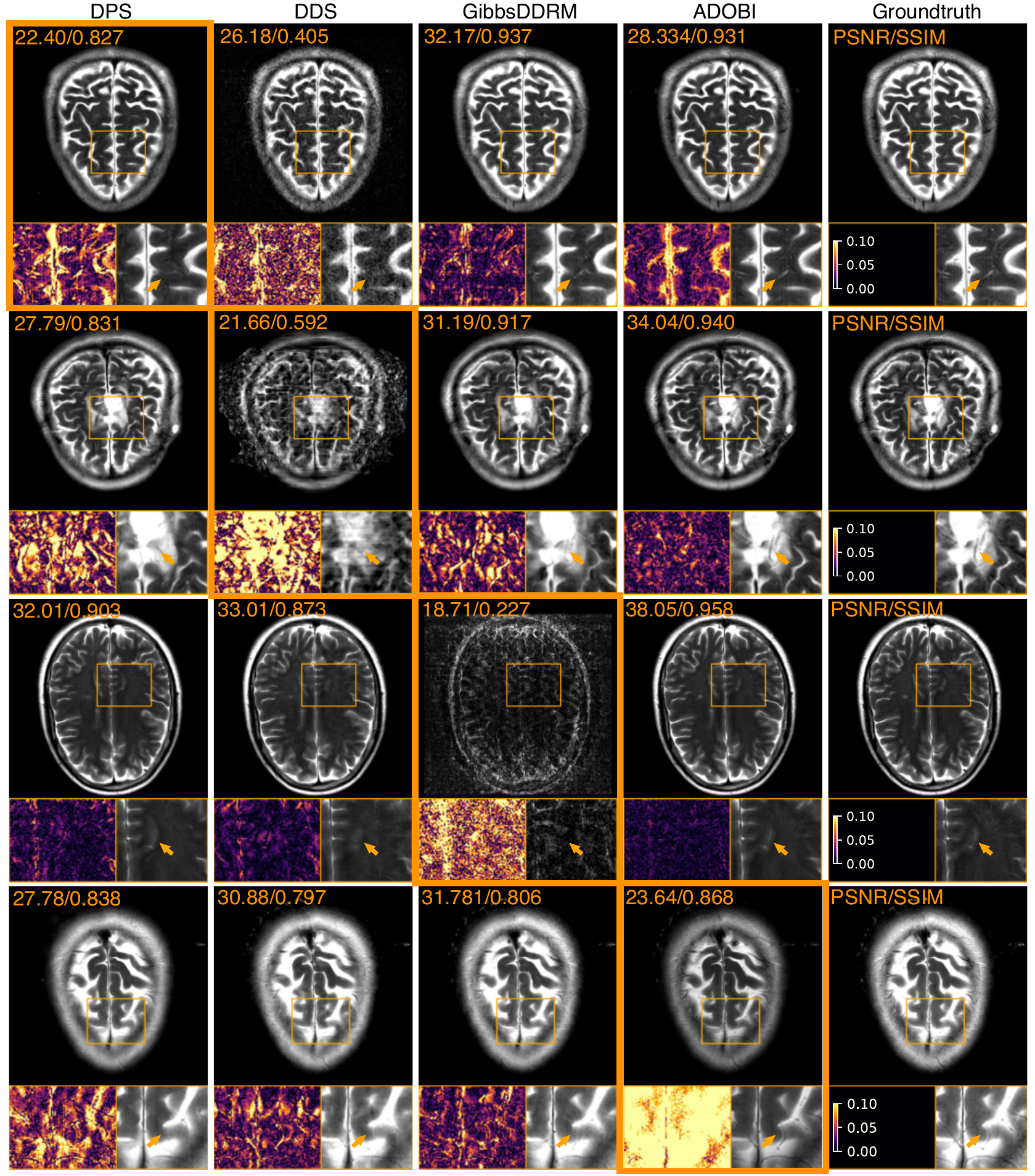}
    \caption{Worst case comparison on 8x acceleration MRI images. The rows from top to bottom denote the lowest-PSNR slices for DPS, DDS, GibbsDDRM, and ADOBI, respectively. We box the particular worst-case result in orange, and display corresponding results from other methods on the same slice for each row.}
        \label{fig:worst_case_8x}
\end{figure*}

\begin{figure*}[h]
\centering    \includegraphics[width=\linewidth]{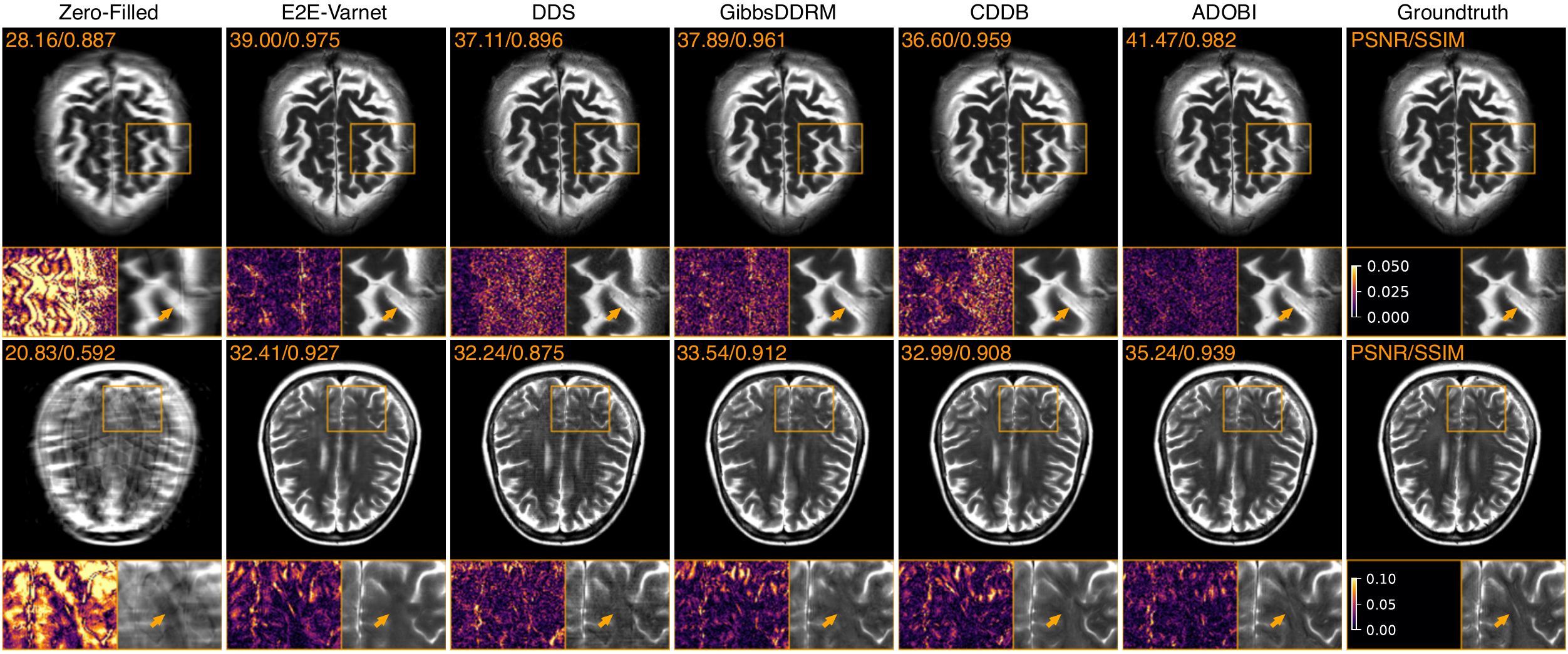}
        \caption{Visual illustration of ADOBI for MRI reconstruction compared to several baseline methods under \textit{noisy} settings. The top row shows results for $4\times$ accelerated PMRI data collection while the bottom row shows those $8\times$. Error maps and zoomed in details highlight visual differences. Note how ADOBI provides the best visual and quantitative performance in both settings.}
        \label{fig:}
\end{figure*}

\end{document}